\documentstyle[psfig,nicV]{article}
\begin{document}
%
%
\heading{%
LiBeB and the Origin of the Cosmic Rays \\
%
}
\par\medskip\noindent
%
\author{%
Reuven Ramaty$^{1}$, Richard E. Lingenfelter$^{2}$, and Benzion 
Kozlovsky$^{3}$
}
\address{
Goddard Space Flight Center, Greenbelt, MD 20771, USA
}
\address{
University of California San Diego, LaJolla, CA 92093, USA
}
\address{
School of Physics and Astronomy, Tel Aviv University, Israel
}

\begin{abstract} Be abundances at low metallicities have major 
implications on cosmic ray origin, requiring  acceleration out of 
fresh supernova ejecta. The observed, essentially constant Be/Fe 
fixes the Be production per SNII, allowing the determination of the 
energy supplied to cosmic rays per SNII. The results rule out 
acceleration out of the metal-poor ISM, and favor Be production at 
all epochs of Galactic evolution by cosmic rays having the same 
spectrum and source composition as those at the current epoch. 
Individual supernova acceleration of its own nucleosynthetic 
products and the collective acceleration by SN shocks of 
ejecta-enriched matter in the interiors of superbubbles have been 
proposed for such origin. A new Monte-Carlo based evolutionary code, 
into which the concept of Be and Fe production per supernova is 
explicitly built in, yields the supernova acceleration efficiency, 
2\% for the refractory metals and 12\% for all the cosmic rays. 
\end{abstract}

\section{Introduction and Overview}

It is well known \cite{reeves94} that cosmic ray interactions have 
an important role in producing the Galactic inventories of Li, Be 
and B (hereafter LiBeB), in particular Be, $^6$Li and $^{10}$B which 
are almost certainly produced solely in cosmic ray interactions. Up 
until recently LiBeB research was dominated by the standard Galactic 
cosmic ray (hereafter sGCR) paradigm which posits that the current 
epoch cosmic rays are accelerated out of an ambient medium of solar 
composition, and that at all past epochs the composition of the 
source particles of the cosmic rays was scaled to solar using the 
ISM metallicity at that epoch. The excess of the observed Be 
abundances in low metallicity stars over the sGCR prediction was 
discussed \cite{pagel91} as early as 1991, the focus of the 
discussion being on whether or not the excess was due to 
contributions from Big Bang nucleosynthesis. But recent calculations 
\cite{orito97} show that the Big Bang contribution to Be production 
is insignificant in comparison with the available Be data at even 
the lowest metallicities. However, these Be observations have major 
implications on cosmic ray origin.

As additional Be data accumulated (see \cite{vroc98} for a recent 
compilation), it became clear that the dependence of log(Be/H) on 
[Fe/H] is essentially linear, not quadratic, as predicted by the 
sGCR paradigm. (Chemical or isotopic symbol ratios denote abundance 
ratios by number and [Fe/H]=log(Fe/H)-log(Fe/H)$_\odot$.) The 
implications of the linear evolution become more obvious 
\cite{rkl98} when log(Be/Fe), rather than log(Be/H) is considered 
(Fig.~1a). Here the horizontal line provides \cite{vroc98} the best 
fit to the data for [Fe/H]$<-1$. Fe production in this epoch 
\cite{timmes95} is dominated by core-collapse supernovae (hereafter 
SNII), with an IMF averaged Fe yield per SNII of $\sim$0.1M$_\odot$, 
essentially independent of metallicity \cite{ww95}. This constancy, 
coupled with the observed constant Be/Fe, requires that, on average, 
about $0.1\times 1.45\times 10^{-6}\times (9/56) = 2.3\times10^{-8} 
{\rm M_\odot}$ of beryllium be produced per SNII, independent of 
[Fe/H]. Under the sGCR paradigm, the cosmic ray composition 
(particularly C/H and O/H) would evolve proportionally to that of 
the ISM, and the Be yield per supernova would increase with [Fe/H], 
contrary to the requirement of the data.

\begin{figure}
\centerline
{\vbox{\psfig{figure=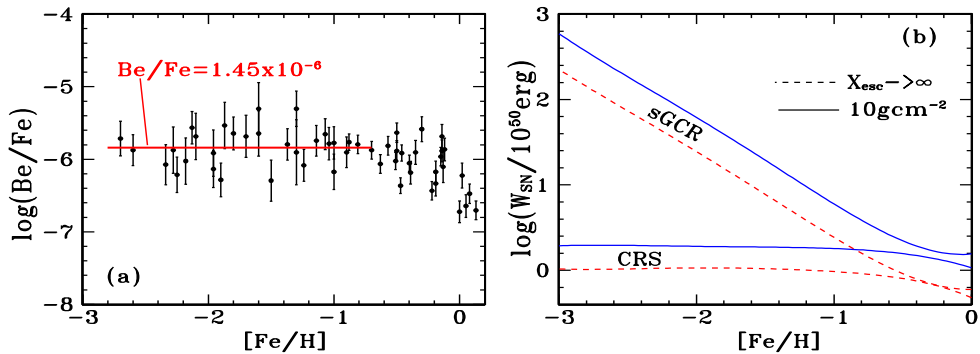,height=4.7cm,width=\textwidth}}}
\caption[]{\small (a) - Observed Be-Fe abundance ratio as a function 
of [Fe/H]; data compilation \cite{vroc98}. The  best fit, for 
[Fe/H]$\lsim$-1, implies that 2.3$\times$10$^{-8}$M$_\odot$ of Be 
are produced per average SNII. The decrease at higher [Fe/H] is due 
to contributions of Type Ia supernovae which make Fe but very 
little Be. (b) - Energy in cosmic rays per SNII required to produce 
this Be mass. The cosmic ray source composition is metallicity 
independent for the CRS model and metallicity scaled for the sGCR 
model. X$_{\rm esc}$$\simeq10$ g cm$^{-2}$ is the current epoch 
cosmic ray escape length; in the early Galaxy it could have been 
different, depending on the density and magnetic structure of the 
early Galactic halo. When X$_{\rm esc}$$\rightarrow$$\infty$, the 
cosmic rays are trapped in the halo until they are either stopped by 
Coulomb collisions or destroyed by nuclear reactions; this choice of 
escape length yields the lowest $W_{\rm SN}$ for the given  Be 
production.} \end{figure}

One solution to this problem was suggested \cite{casse95}, following 
the reported \cite{bloemen94} detection with COMPTEL/CGRO, of C and 
O nuclear gamma ray lines from the Orion star formation region. 
These gamma rays were attributed to a postulated low energy cosmic 
ray (LECR) component, and it was suggested that such LECRs might be 
responsible for the Be (and B) production in the early Galaxy. The 
motivation for this idea (hereafter the LECR model) was the 
indication (based on the reported spectrum of the line emission) 
that the LECRs in Orion are enriched in C and O relative to protons 
and $\alpha$ particles (see \cite{ramaty96} for review and 
\cite{rkl96} for extensive calculations of LiBeB production by 
LECRs). It was suggested \cite{bykov95,rkl96,parizot97} that such 
enriched LECRs might be accelerated out of metal winds of massive 
stars and the ejecta of supernovae from massive star progenitors 
($>$60M$_\odot$) which explode within the bubble around the star 
formation region due to their very short lifetimes. These arguments 
led to the suggestion \cite{vf96,vf97} that the composition of the 
LECRs could be independent of Galactic metallicity, thus allowing 
them to reproduce the linear evolution of Be and B in the early 
Galaxy. The problem with this model is its energetic inefficiency, 
as we shall see. In addition, the validity of the Orion gamma ray 
line observations has been questioned by the COMPTEL team (private 
communication, V. Sch\"onfelder, 1998). 

In a series of publications, co-authored with H. Reeves 
\cite{rrlk97,rklr98,rklr97}, we calculated the energy, $W_{\rm 
SNII}$, in cosmic rays per core collapse supernova, needed to 
produce the required amount of Be. $W_{\rm SNII}$ depends on the 
composition of both the ISM and the cosmic rays, on the energy 
spectrum of the cosmic rays, and on the cosmic ray escape length 
from the Galaxy, X$_{\rm esc}$ measured in g cm$^{-2}$. In Fig.~1b 
we show $W_{\rm SNII}$, as a function of [Fe/H] for two values of 
X$_{\rm esc}$, and for two Galactic cosmic ray origin models: a 
proposed CRS model for which the cosmic rays at all [Fe/H] have the 
same source composition and spectrum as the current epoch cosmic 
rays, and the sGCR model for which the cosmic rays are accelerated 
out of a metallicity dependent ISM with an energy spectrum that is 
also identical to that of the current epoch cosmic rays. The ambient 
ISM composition for both models is solar, scaled with 10$^{\rm 
[Fe/H]}$, except that we allowed O/H to increase by 0.5 dex for 
[Fe/H]$<-1$ (see \cite{pagel97}, figure 8.5). We also increased the 
sGCR cosmic ray C/H and O/H relative to the corresponding 
metallicity dependent ISM values by factors of 1.5 and 2, 
respectively, consistent with the abundances expected for the shock 
accelerated ISM \cite{ellison97}. We see that the sGCR model 
requires that the cosmic ray energy supply per SNII not only be 
metallicity dependent, which is unlikely, but also untenably large, 
exceeding the total available ejecta kinetic energy 
($\sim$1.5$\times$10$^{51}$erg) when [Fe/H]$<-2$. This reinforces 
the previous conclusion that cosmic rays accelerated out of the 
average, metal poor ISM cannot be responsible for the Be production 
in the early Galaxy.

For the CRS model, on the other hand, $W_{\rm SNII}$ is essentially 
constant (Fig.~1b), equal to the very reasonable value of 
$\sim$10$^{50}$ erg/SNII, practically the same as the energy 
supplied per supernova to the current epoch cosmic rays 
\cite{lingen92}. That these two energies are consistent, led to a 
different cosmic ray paradigm, the direct acceleration out of fresh 
supernova ejecta, at least for the refractory metals 
\cite{rkl98,lrk98,hlr98}. The linearity of the Be evolution is a 
straightforward consequence of such a model. To account for the 
various features of the cosmic rays, two related CRS scenarios were 
developed. The first considers the individual supernova shock 
acceleration of its own nucleosynthetic products \cite{lrk98}, while 
the second addresses the collective acceleration by successive SN 
shocks of ejecta-enriched matter in the interiors of superbubbles 
\cite{hlr98}. In the individual supernova model, freshly formed high 
velocity grains in the slowing ejecta reach the forward supernova 
shock which then accelerates the grain erosion products. The 
superbubble model emphasizes the fact that the bulk of the SNIIs 
occur in the cores of superbubbles where the ambient matter is 
likely to be dominated by fresh supernova ejecta. In both scenarios, 
grain erosion products play a central role. They provide an 
explanation for the observed cosmic ray enrichment of the highly 
refractory Mg, Al, Si, Ca, Fe and Ni relative to the highly volatile 
H, He, N, Ne, Ar, an idea developed in detail previously for sGCR 
acceleration \cite{meyer97}. In both the individual supernova and 
superbubble scenarios, the accelerated C and O originate from 
grains, O from oxides (MgSiO$_3$, Fe$_3$O$_4$,Al$_2$O$_3$,CaO) and C 
mainly from graphite. The CRS model based on acceleration within 
superbubbles differs from the LECR model discussed above in that it 
relies on essentially all the SNIIs, and not just those originating 
from massive stellar progenitors, and because it accelerates cosmic 
rays with a spectrum extending to relativistic energies, rather than 
just to low energies. It is these differences that avoid the 
energetic difficulties of the LECR model. Furthermore, while both 
the individual supernova and the superbubble scenarios address the 
origin of the current epoch, well observed cosmic rays, there is no 
hard evidence for the existence of a separate LECR component. Such 
evidence could come from the discovery of Galaxy-wide nuclear 
deexcitation gamma ray line emission (see \cite{rkl79,bloemen97}). 

The B isotopic ratio, $^{11}$B/$^{10}$B= 4.05$\pm$0.2 in meteorites 
\cite{ch95} and 3.4$\pm$0.7 in the current epoch ISM 
\cite{lambert98}, has major implications on the origin of this 
element. These ratios exceed the predictions of both the CRS and 
sGCR models. LECRs could account for these data, either if the 
accelerated particles were confined to very low energies ($\sim$10 
MeV/nucleon), which is energetically untenable, or if their spectrum 
is artificially chosen to yield nuclear interactions predominantly 
at $\sim$30 MeV/nucleon, which is also improbable (see 
\cite{rklr97}). The excess $^{11}$B is most likely the result of 
$^{12}$C spallation by neutrinos in SNIIs \cite{woos90}. 

Concerning Li, it is clear that the bulk of the $^7$Li at all epochs 
of Galactic evolution did not originate in cosmic ray interactions. 
On the other hand, all of the $^6$Li is most likely cosmic ray 
produced. The CRS model predicts that $^6$Li/Be is about 5 at all 
metallicities, depending somewhat on the uncertainties in the 
$\alpha$$\alpha$ cross section at high energies (see \cite{rklr97}). 
This value is quite consistent with the meteoritic ratio of 5.8 
\cite{ag89}. At low metallicities, higher values of 
$^6$Li/Be were reported \cite{hobbs97}, but the uncertainties in 
these data are very large.

To explore these problems, we have developed a Monte-Carlo based 
evolutionary code into which the concept of Be and Fe production per 
supernova is explicitly built in. The treatment allows the easy 
investigation of a variety of hitherto ignored processes, such as 
the delay in Be production relative to Fe production due to the 
propagation of the cosmic rays, and the dependencies of the cosmic 
ray composition and ejecta kinetic energy on supernova progenitor 
mass.

\begin{figure}
\centerline
{\vbox{\psfig{figure=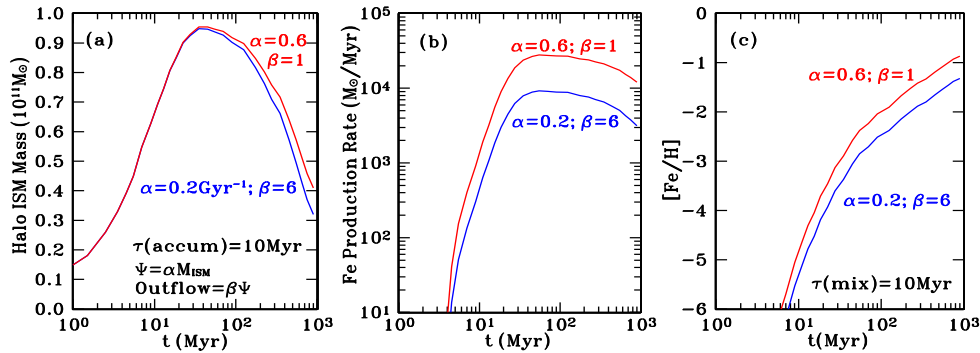,height=4.7cm,width=\textwidth}}}
\caption[]{\small (a) -  Evolution of the halo ISM mass, M$_{\rm 
ISM}$, accumulating at a rate [10$^{11}$M$_\odot$/$\tau({\rm 
accum})]$e$^{-t/\tau({\rm accum}})$, where $\tau({\rm accum})$=10Myr 
is not varied as it does not significantly affect the relevant 
abundance ratios. $\Psi$=$\alpha$M$_{\rm ISM}$ and $\beta$ $\Psi$ 
are the mass removal rates by star formation and outflow to the 
disk, respectively; we take $\alpha$=0.2Gyr$^{-1}$, $\beta$=6 (see 
\cite{prantzos93}) and $\alpha$=0.6Gyr$^{-1}$, $\beta$=1, an example 
of faster star formation but slower outflow. (b) and (c) - Time 
dependent Fe production at the time of nucleosynthesis and halo ISM 
Fe abundance  taking into account a delay due to transport and 
mixing (see text). }
\end{figure}

\section{Evolution}

We consider a one-zone model, representing the halo of the early 
Galaxy, and limit our treatment to the first 10$^9$ years of 
Galactic evolution, as this low metallicity ([Fe/H]$\lsim$$-1$) era 
conveys the most unambiguous information on cosmic rays origin. We 
assume that the halo ISM mass accumulates exponentially, and we 
remove mass by star formation and by outflow to the disk. In the 
Monte Carlo simulation we set up a grid of time bins, and for each 
bin generate an ensemble of stellar masses using the Salpeter IMF. 
Starting from the first bin and advancing in time, we calculate the 
removed mass. Then depending on stellar lifetimes (see 
\cite{pagel97} table 7.1) and a widely used expression for the 
remnant mass (e.g. \cite{prantzos93}), we calculate the returned 
mass for each star in the ensemble, but note that only $>$2M$_\odot$ 
stars return mass for $t$$<$1 Gyr. The resultant time dependence of 
M$_{\rm ISM}$ is shown in Fig.~2a. 

Fe production is due to the $>$10 M$_\odot$ stars which explode as 
SNIIs. We calculate the Fe mass and production time for every such 
star using SNII yields \cite{ww95} at zero metallicity and stellar 
lifetimes. The Fe production rates are shown in Fig.~2b for the two 
choices of mass loss. The production rate goes through a maximum 
between a few tens and a hundred Myr. As the average Fe yield per 
SNII is $\sim$0.1 M$_\odot$, the corresponding maximum SNII rate is 
10-30/century.  We calculate the Fe content of the ISM by delaying, 
due to transport and mixing, the deposition of the synthesized Fe. 
In the simulation we choose the mixing time randomly in the interval 
0 to $\tau({\rm mix})$. The resultant Fe abundance evolution is shown 
in Fig.~2c for $\tau({\rm mix})$=10 Myr. We see that [Fe/H] is 
indeed $\sim -1$ at 1 Gyr, with a somewhat larger value for the 
case of larger star formation and lower outflow rate. As we shall 
see, variations in $\tau({\rm mix})$ can have significant effects on 
Be/Fe, B/Be and $^{11}$B/$^{10}$B at very low [Fe/H].

\begin{figure}
\centerline
{\vbox{\psfig{figure=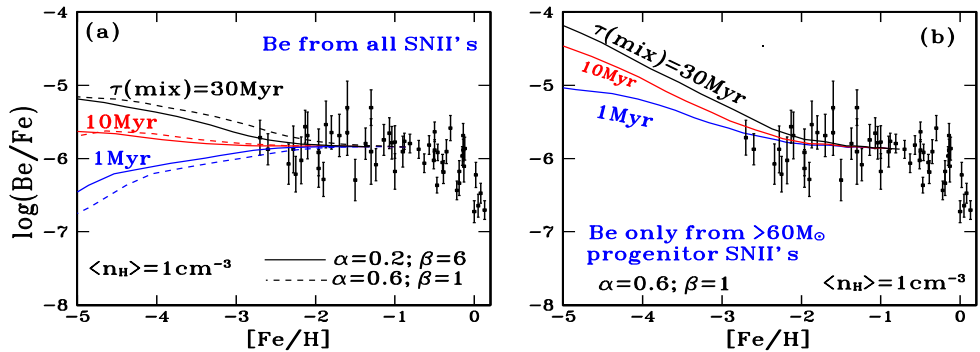,height=4.7cm,width=\textwidth}}}
\caption[]{\small (a) - The evolution of Be/Fe. (a) - the CRS model; 
(b) - the LECR model. For $-2.5\lsim$[Fe/H]$\lsim-1$ the only free 
parameter is the refractory cosmic ray acceleration efficiency, 
$\eta_{\rm refr}$, equal to 0.018 and 0.63 for the two models, 
respectively. The effects seen at lower [Fe/H] are discussed in the 
text.}
\end{figure}

We carry out a similar procedure for Be. Here we distinguish two 
cases which are analogous to the CRS and LECR models discussed 
above. In the first one (CRS), we assign to each $>$10 M$_\odot$ 
star an ejecta kinetic energy $W_{\rm ej}$, using the values given 
in \cite{ww95}. We then assume that each corresponding SNII imparts 
$\eta_{\rm refr}$$W_{\rm ej}$ ergs to refractory cosmic rays and 
take $\eta_{\rm refr}$ to be independent of progenitor mass. Next we 
calculate the relative abundances of the refractories 
(C:O:Mg:Al:Si:Ca:Fe:Ni) as a function of progenitor mass, using the 
ejected masses from \cite{ww95} and assuming \cite{lrk98} that the O 
abundance is determined by the amount of O in oxides (MgSiO$_3$, 
Fe$_3$O$_4$,Al$_2$O$_3$,CaO) and that the fraction of C that 
condenses in grains is the same as that of the highly refractories 
Mg, Al, Si and Fe. Then by employing the LiBeB production code 
developed previously \cite{rklr97}, with $E_0$=10 GeV/nucleon 
appropriate for high energy cosmic rays, we calculate the total Be 
mass produced by each SNII in the simulation. For the second case 
(LECR), we follow a similar procedure, except that we only consider 
$>$60 M$_\odot$ progenitors and take $E_0$=30 MeV/nucleon. The only 
free parameter in these calculations is $\eta_{\rm refr}$. The time 
interval between cosmic rays acceleration and Be production can be 
quite large due to cosmic ray transport. Therefore, for both 
cases we delay the Be production relative to that of the Fe, 
employing randomly distributed delay times in the range 0 - 10 Myr 
(CRS) and 0 - 1 Myr (LECR), valid for an ISM density of 1 H 
cm$^{-3}$. The delay for LECRs is smaller because they interact on 
shorter time scales than their high energy counterparts. 

Fig.~3 shows log(Be/Fe) vs. [Fe/H], where the normalization to the 
data yields the fraction of supernova kinetic energy going into 
refractory cosmic rays, $\eta_{\rm refr}$ = 0.018 and 0.63 for the 
CRS and LECR cases, respectively. The much larger LECR value is due 
to the less frequent $>$60 M$_\odot$ progenitors explosions and to 
the less efficient Be production by LECRs \cite{rklr97}. The 
acceleration of volatile elements (in particular protons and 
$\alpha$ particles) that should accompany the acceleration of the 
metals requires more energy. For the current cosmic ray source 
composition, the ratio of the total energy in cosmic rays per SNII 
and the SNII ejecta energy is 0.12 for the CRS case and 4.3 for the 
LECR case. While this CRS value is very reasonable, the LECR value 
is clearly excessive and would invalidate the model, unless the 
SNIIs of very massive progenitors only accelerate metals. The 
energetic efficiency of the LECR case could also be improved by 
choosing a harder accelerated particle spectrum (e.g. $E_0 \simeq 
70$ MeV/nucleon \cite{rklr97}) and a progenitor mass cutoff smaller 
than 60 M$_\odot$.   

The behavior of log(Be/Fe) at [Fe/H]$<-3$ shows interesting effects. 
An increase in Be/Fe, caused by the short lifetimes of the very 
massive progenitors, was predicted for an LECR case previously 
\cite{vroc98}. The $\tau({\rm mix})$=1 Myr curve in Fig.~3b shows 
that predicted evolution. The introduction of a mixing time for Fe 
enhances the effect. On the other hand, for the CRS case (Fig.~3a) 
where the Be production is delayed, the effect can be canceled or 
even reversed, depending on the mixing time relative to the Be 
production time.

\begin{figure}
\centerline
{\vbox{\psfig{figure=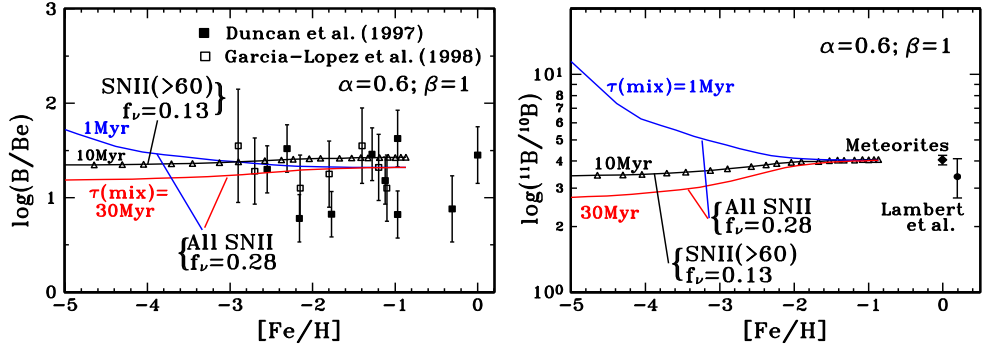,height=4.7cm,width=\textwidth}}}
\caption[]{\small (a) - Evolution of B/Be; data from 
\cite{duncan97} and  \cite{garcia98}. (b) - 
evolution of $^{11}$B/$^{10}$B - data from \cite{ch95} and 
\cite{lambert98}.
}
\end{figure}

We treat the evolutions of $^{10}$B and the cosmic ray produced 
$^{11}$B in the same fashion as the Be evolution. For the  
$\nu$-produced $^{11}$B we use the SNII yields \cite{ww95} at zero 
metallicity scaled by a factor $f_\nu$, and treat the evolution 
similar to that of the Fe. We determined $f_\nu$ by requiring that 
when the effects of mixing and cosmic ray production times become 
negligible ([Fe/H]$\gsim-2$, Fig.~4b), $^{11}$B/$^{10}$B = 4, the 
meteoritic value \cite{ch95} at [Fe/H]=0. We thus obtain 
$f_\nu$=0.28 (CRS) and 0.13 (LECR). In a past calculation 
\cite{vf96}, $^{11}$B/$^{10}$B was shown to be $\sim$7 at 
[Fe/H]$\simeq-2$ and to decrease toward 4 at higher metallicities. 
This resulted from the added contribution of the sGCR at 
[Fe/H]$>-1$. But we suspect that the increase \cite{ww95}, by a 
factor of $\sim$2, of the $\nu$-produced $^{11}$B per SNII with 
increasing [Fe/H], was not taken into account in that calculation, 
an effect that would have canceled the decrease and would have 
allowed $^{11}$B/$^{10}$B$\simeq$4 at [Fe/H]$\simeq-2$, as in our 
calculation. For the CRS case (Fig.~4a) the calculated B/Be for 
$-3$$<$[Fe/H]$<$$-1$ is 17-21 and 24-21 for $\tau({\rm mix})$=30Myr 
and 1Myr, respectively. These are in good agreement with the 
observations (e.g. B/Be=20$\pm$10 \cite{garcia98}). For the LECR 
case and the same [Fe/H] range, B/Be is 24-26, also in agreement 
with the data. We note that the total B in both the CRS and LECR 
cases includes $\nu$ production. For [Fe/H]$<$$-3$, both B/Be and 
$^{11}$B/$^{10}$B exhibit interesting effects. The very large 
$^{11}$B/$^{10}$B (Fig.~4b) is the consequence of delayed cosmic ray 
production and possible, prompt $\nu$-induced production in SNIIs.

\vfill
\end{document}